\documentclass{sig-alternate-2013}

\permission{Copyright is held by the author/owner(s).}
\conferenceinfo{WWW'16 Companion,}{April 11--15, 2016, Montr\'eal, Qu\'ebec, Canada.} 
\copyrightetc{ACM \the\acmcopyr}
\crdata{978-1-4503-4144-8/16/04. \\
http://dx.doi.org/10.1145/2872518.2890549 }

\begin{document}






%

\title{Search by Ideal Candidates: Next Generation of Talent Search at LinkedIn}

%
%
%
%
%

\numberofauthors{1} 
%

\author{
%
%
\alignauthor
{Viet Ha-Thuc, Ye Xu, Satya Pradeep Kanduri, Xianren~Wu, Vijay~Dialani, Yan~Yan, Abhishek~Gupta, Shakti~Sinha\titlenote{All authors equally contribute to the paper.}}\\
\affaddr{LinkedIn Corporation\\ 2029 Steirlin Ct, Mountain View, CA 95035, USA}\\
\email{\{vhathuc,yexu,skanduri,rwu,vdialani,cyan,agupta,ssinha\}@linkedin.com}
}


\maketitle
\begin{abstract}
One key challenge in talent search is how to translate complex criteria of a hiring position into a search query. This typically requires deep knowledge on which skills are typically needed for the position, what are their alternatives, which companies are likely to have such candidates, etc.
However, listing examples of suitable candidates for a given position is a relatively easy job. 
Therefore, in order to help searchers overcome this challenge, we design a next generation of talent search paradigm at LinkedIn: Search by Ideal Candidates. This new system only needs the searcher to input one or several examples of suitable candidates for the position. The system will generate a query based on the input candidates and then retrieve and rank results based on the query as well as the input candidates. The query is also shown to the searcher to make the system transparent and to allow the searcher to interact with it. As the searcher modifies the initial query and makes it deviate from the ideal candidates, the search ranking function dynamically adjusts an refreshes the ranking results balancing between the roles of query and ideal candidates. As of writing this paper, the new system is being launched to our customers.
\end{abstract}

%
%


%
%

%
%



\section{Introduction}
LinkedIn is the world largest professional networking platform with more than 400 million members.
Within every second,
more than two new members join the website. Our vision is to become the first economic graph matching talents and opportunities at a massive scale. Over the time, as our member base increases and our economic graph vision becomes reality, LinkedIn has become the source for corporations around the world to find new hires. On financial side, about {\it 64\%} of the company revenue is from Talent Solutions\footnote{http://blog.linkedin.com/2015/10/29/linkedins-q3-2015-earnings/}, which is a product helping recruiters and corporations around the world search for suitable talents. Therefore, talent search problem is extremely important to LinkedIn. 

A key challenge in talent search is to translate the criteria of a hiring position into a search query that leads to desired candidates. To fulfill this goal, the searcher has to understand which skills are typically required for the position, what are the alternatives, which companies are likely to have such candidates, which schools the candidates are most likely to graduate from, etc. Moreover, the knowledge varies overtime. As a result, it is not surprising that even for experienced recruiters, it often requires many searching trials in order to get to a good query, as indicated by LinkedIn search log data. 

On the other hand, it is usually quite straightforward for a searcher to list one or several examples of good candidates for a given position. For instance, hiring managers or recruiters can simply source from the existing team members. Motivated by this, we propose a new talent search paradigm called \textit{Search by Ideal Candidates}. In this paradigm, instead of specifying a complex query capturing the position requirements, the searcher can simply pick up a small set of ideal candidates (typically from one to three candidates) for the position. The system then builds a query automatically extracted from the input candidates and searches for result candidates who are similar to them. It is worth emphasizing that the query built from the ideal candidates is not only an intermediate goal used to find result candidates but also shown to the searcher. Exhibiting the query explains why a certain result is shown up in search ranking thus makes the system transparent to the searcher. Furthermore, it allows the searcher to interact with the system and have control over the results by modifying the initial query. Therefore, building a high quality (initial) query and retrieving relevant results are equally important goals in this search paradigm. 

Our system is related to several previous research directions in the literature including query-by-example typically used in image retrieval \cite{ballerini2010query}, relevance feedback popularly used in text retrieval \cite{rocchio1971relevance} and item-to-item recommendation \cite{sarwar2001item}. However, our problem has unique challenges. First, unlike the recommendation problem, in Search by Ideal Candidates, it is not only important to generate relevant results but also critical to build descriptive queries as explained above. Second, LinkedIn data is semi-structured and has unique characteristics. Different attribute types like skill, job title, company, etc. require different query building strategies. Third, as the searcher modifies the query, his information requirement could start slightly shifting away from the initial candidate inputs. Thus, the roles of the query and ideal candidates on results dynamically change within a session. 

To overcome these challenges, for query building, we propose an approach exploiting the uniqueness of LinkedIn data to generate queries from ideal candidates. In particular, to generate skills representing the expertise of the candidates, our approach estimates expertise scores of each candidate on the skills he or she might have and then aggregates the scores across the candidates to determine the most representative ones. To generate company attribute, our approach exploits co-viewing relationships amongst companies to discover companies similar to input candidates' companies. To rank the final results, we propose a ranking function taking both query and input candidates into account. As the query increasingly deviates from the input candidates, the ranking function gradually focuses more on the impact of the query than it of the input candidates.

The new talent search paradigm is expected to be the next generation of LinkedIn Talent Solution. As of this writing, the product is being launched to a few pilot customers. We plan to ramp it up to other external customers in early 2016.

\section{System Overview}
\begin{figure}
\centering
\includegraphics[width=0.4\textwidth]{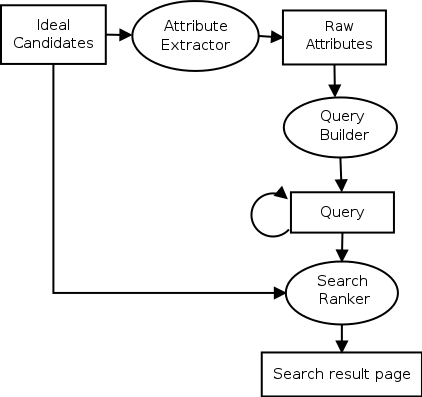}
\caption{System Overview}
\label{system_overview}
\end{figure}

Given a set of input candidates, our goal is to build a search query capturing the key information in their profiles and use the query to retrieve and rank results. The overall flow is shown in Figure \ref{system_overview}. In the first step, we extract the raw attributes, including skills, companies, titles, schools, industries, etc. from their profiles individually. These raw attributes are then passed to a query builder. For each attribute type, the query builder aggregates the raw attributes across the input candidates, expands them to similar attributes and finally selects the top attributes that are most similar to the ones from the input candidates. 

After the query is generated, it is shown to the searcher and used to search for results. The searcher can interact and modify the query, e.g., adding or removing some attributes like skills or companies in the query. After the query is modified, search ranker will refresh search results. The search ranker takes both the query (initial or modified one) as well as the input candidates into account when ranking search results. In the next sections, we discuss in more details the two key components of the system: query building and search ranking.

\section{Query Building}
Given the raw attributes from the profiles of the ideal candidates, query builder generates a query containing skills, companies, titles, etc. that best represents the ideal candidates. Given space limit, in this paper we focus on skill and company attributes. 

\subsection{Skills}
LinkedIn allows members to add  {\it skills} to their profiles. Typical examples of skills for an information retrieval researcher would be ``search'', ``information retrieval'', ``machine learning'', etc. On LinkedIn, there are more than 35 thousand standardized skills. Members can also {\it endorse} skills of other members in their network. Thus, skills are an integral part of members' profiles that showcases their professional expertise. A challenge is that the ideal candidates may not explicitly list all the skills they have on profiles. On the other hand, some of their skills might not be relevant to their core expertise. For instance, an information retrieval researcher could have ``nonprofit fundraising'' skill.

\begin{figure}
\centering
\includegraphics[width=0.4\textwidth]{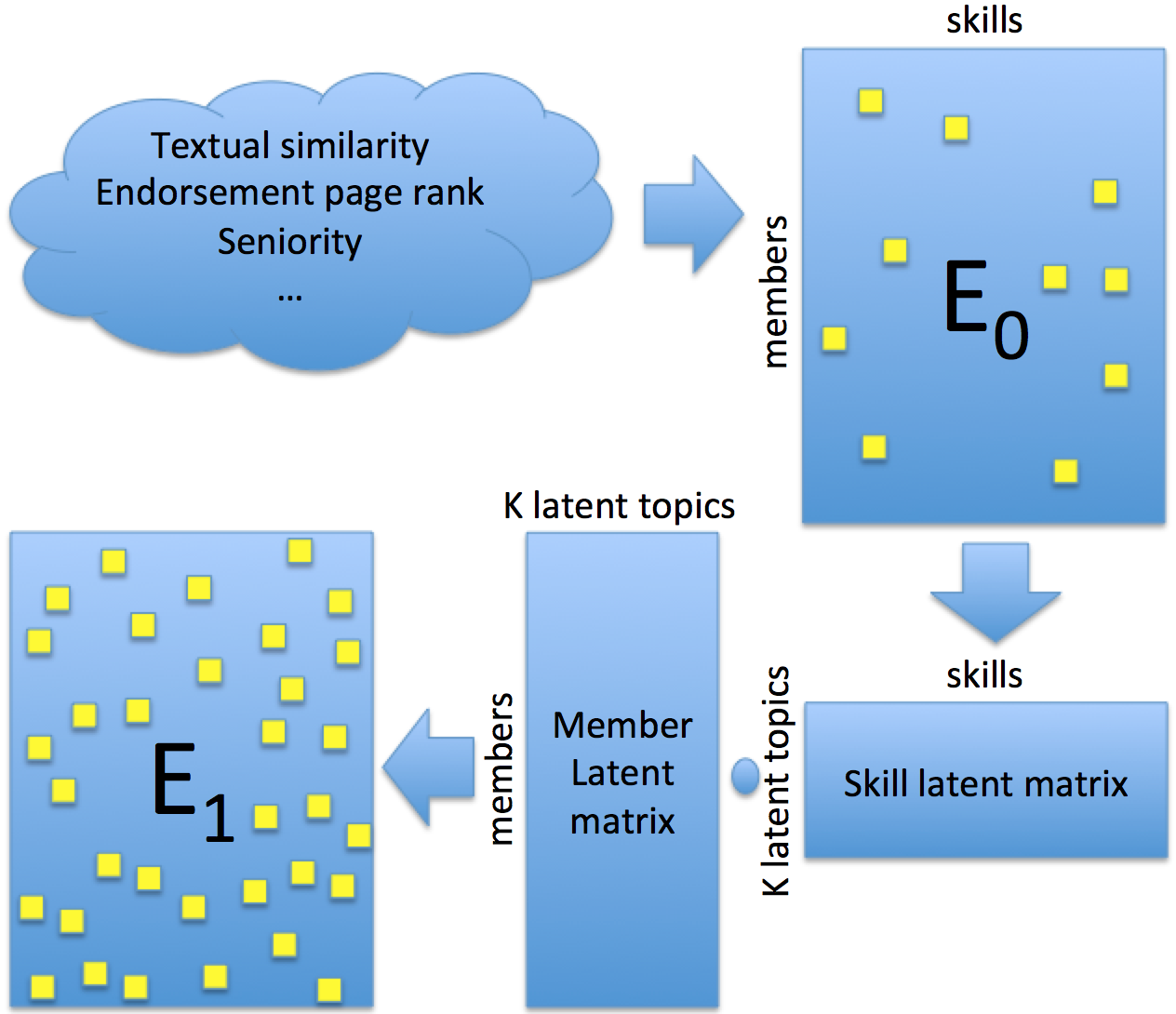}
\caption{Member-skill expertise estimation}
\label{member_expertise}
\end{figure}

To overcome these challenges, we first estimate expertise scores of a member on the explicit skills and the ones he might have. Figure \ref{member_expertise} describes the offline process to estimate the expertise scores. In the first step, we use a supervised learning algorithm combining various signals on LinkedIn such as skill-endorsement graph page rank, skill-profile textual similarity, member's seniority, etc. to estimate the expertise score, i.e., $p(expert | member, skill)$. After this step, the expertise matrix ($E_0$) is very sparse since we can be certain only for a small percentage of the pairs. In the second step, we factorize the matrix into member and skill matrices in K-dimensional latent space. Then, we compute the dot-product of the matrices to fill in the ``unknown'' cells. The intuition is that if a member have ``machine learning'' and ``information retrieval'' skills, based on skill co-occurrence patterns from all of our member base, we could infer that the member is likely to also know ``learning-to-rank''. Since the dot-product results in a large number of non-zero scores of each member on the 35K skills, the scores are then thresholded. If member's score on a skill is less than a threshold, the member is assumed not to know the skill and has zero expertise score on it. Thus, the final expertise matrix ($E_1$) is still sparse but relatively much denser than $E_0$. We refer interested readers to our recent work \cite{hathuc2015expertisesearch} for more details. 

\begin{equation}
	f(skill) = \sum_{c \in IC} {expertiseScore(c, skill)}
\label{skill_ranking}
\end{equation}

At running time, given a set of input ideal candidates $IC$, we rank the skills by the ranking function in Equation \ref{skill_ranking}. The top-N skills are selected to represent the ideal candidates. Expertise scores of an ideal candidate on outlier skills are zero or very low, thus these skills are unlikely to get selected. Moreover, by taking the sum over all candidates, we boost the skills which many candidates are with, thus representing the commonality of the skill set. 
 
\subsection{Companies}
Given the ideal candidates, besides their own companies, the query builder generates a set of companies, which are likely to also have candidates similar to them. In this work, we exploit the uniqueness of LinkedIn data to find such companies.

Similar to our approach to finding missing skills, we again apply collaborative filtering \cite{Wu14} exploiting the ``wisdom of the crowd'' to find company relationships. Specifically, we construct a company browsemap using co-viewing relationship (people who view company A also view company B). Intuitively, companies co-viewed by highly overlapped sets of people are likely to be similar. Given company similarities, the query builder takes the input ideal candidates' current companies as well as top-N similar companies as company facet in the query.   

\subsection{Query Building Demo}
\begin{figure}
\centering
\includegraphics[width=0.5\textwidth]{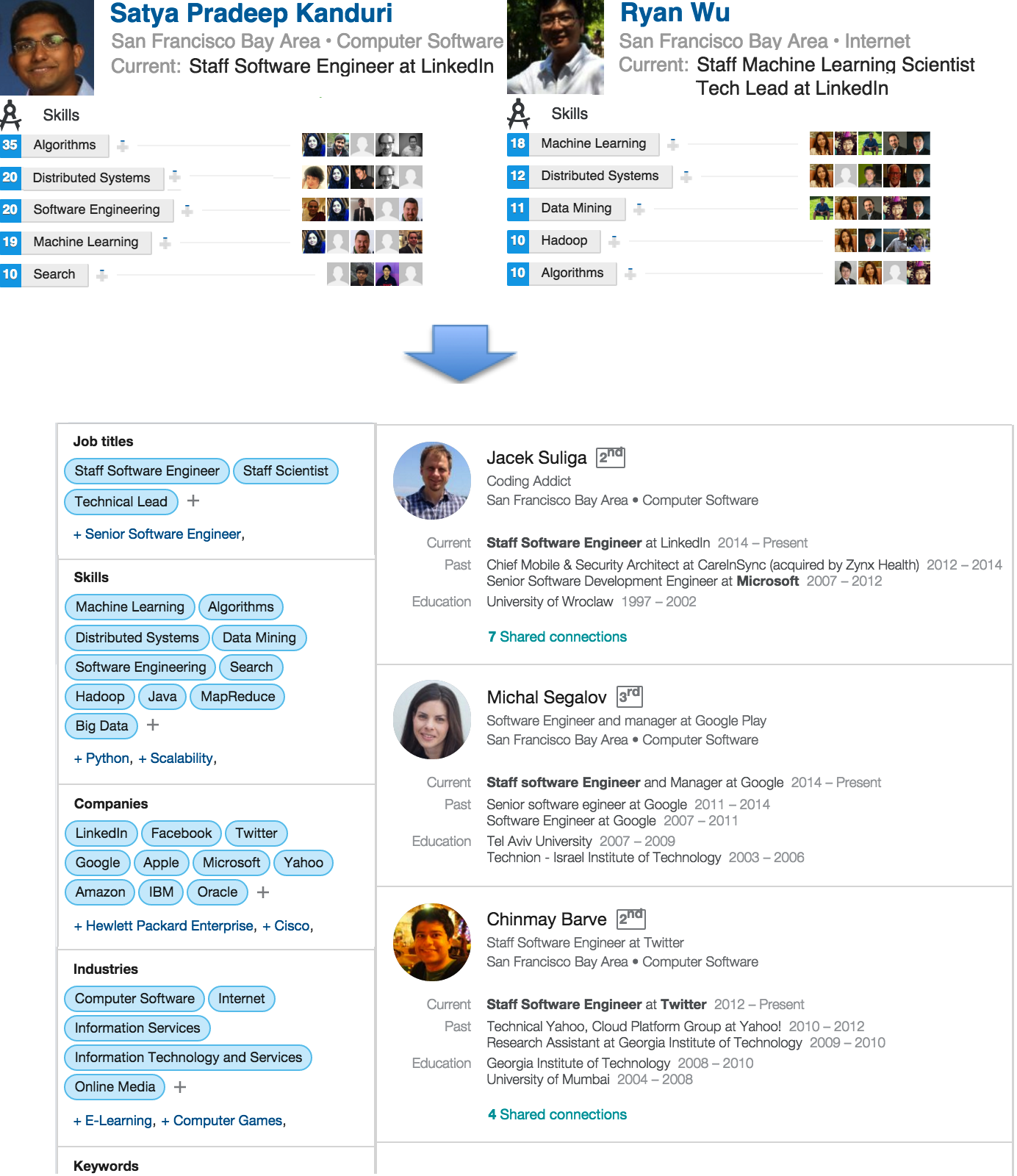}
\caption{Search By Ideal Candidates}
\label{demo}
\end{figure}

Assume our team would like to hire new members, instead of specifying a highly complex query describing the requirement, we can simply take profiles of two team members, Satya Pradeep Kanduri and Ryan Wu, as the input to the search system. The upper part of Figure \ref{demo} shows snippets of their profiles. Given the two ideal candidates, the query builder automatically generates a query as shown on the left rail in the lower part of the figure, including job title, skill, company and industry facets. On skill facet, as shown in the figure, the selected ones represent the core expertise of the ideal candidate. Company facet contains their current company (``LinkedIn'') as well as the ones similar to LinkedIn such as ``Facebook'', ``Twitter'' and ``Google''. These are the companies that are likely to have candidates like Satya and Ryan. Job title facet has \textit{standardized} job titles extracted from the two profiles. In our system, different variations of the same title such as ``tech lead'' and ``technical lead'' are mapped (standardized) to the same title entity. This helps to improve search recall.

When interacting with the query, the searcher can delete any of the entities that he thinks irrelavant. Besides the selected entities in the query, the system also suggests additional ones, e.g., ``python'' and ``scalability'' in skill facet. This would make it easier for the searcher to add new ones. The suggested ones are based on all entities in the current query. When searcher modifies the query, the suggestions are refreshed. Finally, we have \textit{keywords} field that allows the searcher to enter unstructured text to complement the other facets if necessarily.

\section{Search Ranking}
In the next generation of talent search, we propose a new ranking function that takes into account not only query and searcher but also the input ideal candidates (See Equation \ref{ranking_function}). The first part in the function ($f_1$) estimates how a result $r$ is relevant to query $q$ and searcher $s$, as in the standard personalized search. The second part ($f_2$) aims to guarantee a direct similarity between a result and input ideal candidates ($IC$). However, as the searcher keeps modifying the query, the query could increasingly deviate from the original input candidates. Thus, the role of the ideal candidates becomes less important. At the same time, as the search session continues, the fact that the searcher sees attributes and does not remove them increases the confidence that the remaining attributes in the query are useful. Therefore, the importance of the query should increase. To capture this, we apply a decay function ($e^{-\lambda n}$) to control the dynamic balance between the impacts of the query and the input ideal candidates on result ranking. In the decay function, \textit{n} is the number of times the searcher has modified the query and refreshed results. $\lambda$ is a parameter controlling decay rate.     

\begin{equation}
	f(r, q, s, IC) = \frac{f_1(r, q, s)+e^{-\lambda n}f_2(r, IC)}{1+e^{-\lambda n}}
\label{ranking_function}
\end{equation}

\subsection{Personalized Ranking}
The personalized ranking function ($f_1$) combines a variety of signals between a result, query and searcher. Due to business sensitivity, we omit some details regarding to feature computation. Instead, we give a high-level description the key feature categories.

\textbf{Expertise scores} 
For every search result, we compute sum of his or her expertise scores, which are described in Section 2.1, on the skills in the query. The higher the sum is, the better the result matches the information need in terms of skills.

\textbf{Textual features} 
These features capture the textual similarities between the query and different sections of result profiles, such as, current titles, past titles, current companies, past companies, etc.   

\textbf{Geographic features} Talent search on LinkedIn is highly personalized. For instance, given the same
query, recruiters in New York City and in Montreal are interested in very different results. Thus, location plays an important role in personalizing search results. We create multiple features capturing this.

\textbf{Social features} Another important aspect of personalization is to capture how the results socially relate to the searcher. We leverage a variety of the signals on LinkedIn, such as common friends, companies, groups and schools to generate features in this category. 

Interested reader could find more details on how the personalized ranking is trained in our recent work \cite{hathuc2015expertisesearch}. 

\subsection{Career Trajectory Similarity}


Different from keyword based matching methodology, Career Trajectory Similarity (CareerSim) ascertains a similarity between two profiles by leveraging the trajectory information encoded in series of positions held by the individuals through their careers. Due to the space limitation, we focus on the two key steps of the CareerSim framework here.

\textbf{Profile Modeling}
To capture the trajectory information, CareerSim models every individual member profile as a sequence of nodes, each of which records all information within a particular position of member's career, such as company, title, industry, time duration, and keyword summary.

\textbf{Similarity Computing}
At the node (position) level, similarity is ascertained by using a generalized linear model but other approaches could be easily substituted. Then at the sequence (profile) level, 
sequence alignment method is employed to 
find an optimal alignment between
pairs of nodes from the two career paths.

To the best of our knowledge, CareerSim is the first framework to model professional similarity between two people taking into account their career trajectory information. More details about the CareerSim framework can be found in our recent paper \cite{Xu14}. Given a set of ideal candidates $IC$, the similarity between a result $r$ and the candidate set is simply the average over the individual ones as shown in Equation \ref{career_sim}. We posit that using the temporal and structural features of a career trajectory for modeling similarity between a result and the ideal candidates provides a good complement to the signals in the personalized ranker. This also gives a direct similarity between a result and ideal candidates.

\begin{equation}
	f_2(r, IC) = \frac{\sum_{c \in IC} CareerSim(r,c)}{|IC|} 
\label{career_sim}
\end{equation}

\subsection{Search Ranking Demo}

Following the scenario described in Section 3.3, after the query builder constructs a query from input profiles, the query is used to retrieve matched results. Then, the personalized ranker takes the query and the searcher's information into account to score each of the matched results. At the same time, the ideal candidates (Satya and Ryan) are also treated as another input for the CareerSim model to obtain a score measuring career trajectory similarity between each result and the candidates. The final ranking results based on a combination of the two scores are shown in the low-right section of Figure \ref{demo}. As illustrated in the figure, the top results are similar to the ideal candidates. They are all software engineers with the same seniority level (``staff'') and from the same or similar companies (``LinkedIn'', ``Google'' or ``Twitter''). 

During the search section, the searcher is still allowed to interact with the query via deleting or adding any of the entities. After every query edition, the retrieval system and the ranker are triggered and the result ranking is refreshed.

\section{Conclusions}
In this work, we present the next generation of talent search at LinkedIn: Search by Ideal Candidates. In this new search paradigm, instead of constructing a highly complicated query describing criteria of a hiring position, the searcher simply inputs one or a few ideal candidates for the position, e.g., existing members in the team. Our system will automatically build a query from the ideal candidates and then retrieve and rank results.

For query building, we present approaches based on collaborative filtering to generate skill and company facets. For result ranking, we propose a ranking function combining a personalized search ranker and a career trajectory similarity model. As of this writing, the product is being launched to a set of pilot customers and it will be ramped to other customers early 2016. 



%
\bibliographystyle{abbrv}
\bibliography{sigproc}  
%
%
\end{document}